\documentstyle[12pt]{article}
\ifx\undefined\psfig\else\endinput\fi

\edef\psfigRestoreAt{\catcode`@=\number\catcode`@\relax}
\catcode`\@=11\relax
\newwrite\@unused
\def\typeout#1{{\let\protect\string\immediate\write\@unused{#1}}}
\def\figurepath{./}

\def\@nnil{\@nil}
\def\@empty{}
\def\@psdonoop#1\@@#2#3{}
\def\@psdo#1:=#2\do#3{\edef\@psdotmp{#2}\ifx\@psdotmp\@empty \else
    \expandafter\@psdoloop#2,\@nil,\@nil\@@#1{#3}\fi}
\def\@psdoloop#1,#2,#3\@@#4#5{\def#4{#1}\ifx #4\@nnil \else
       #5\def#4{#2}\ifx #4\@nnil \else#5\@ipsdoloop #3\@@#4{#5}\fi\fi}
\def\@ipsdoloop#1,#2\@@#3#4{\def#3{#1}\ifx #3\@nnil 
       \let\@nextwhile=\@psdonoop \else
      #4\relax\let\@nextwhile=\@ipsdoloop\fi\@nextwhile#2\@@#3{#4}}
\def\@tpsdo#1:=#2\do#3{\xdef\@psdotmp{#2}\ifx\@psdotmp\@empty \else
    \@tpsdoloop#2\@nil\@nil\@@#1{#3}\fi}
\def\@tpsdoloop#1#2\@@#3#4{\def#3{#1}\ifx #3\@nnil 
       \let\@nextwhile=\@psdonoop \else
      #4\relax\let\@nextwhile=\@tpsdoloop\fi\@nextwhile#2\@@#3{#4}}
\newread\ps@stream
\newif\ifnot@eof       
\newif\if@noisy        
\newif\if@atend        
\newif\if@psfile       
{\catcode`\%=12\global\gdef\epsf@start{
\def\epsf@PS{PS}
\def\epsf@getbb#1{%
\openin\ps@stream=#1
\ifeof\ps@stream\typeout{Error, File #1 not found}\else
   {\not@eoftrue \chardef\other=12
    \def\do##1{\catcode`##1=\other}\dospecials \catcode`\ =10
    \loop
       \if@psfile
          \read\ps@stream to \epsf@fileline
       \else{
          \obeyspaces
          \read\ps@stream to \epsf@tmp\global\let\epsf@fileline\epsf@tmp}
       \fi
       \ifeof\ps@stream\not@eoffalse\else
       \if@psfile\else
       \expandafter\epsf@test\epsf@fileline:. \\%
       \fi
          \expandafter\epsf@aux\epsf@fileline:. \\%
       \fi
   \ifnot@eof\repeat
   }\closein\ps@stream\fi}%
\long\def\epsf@test#1#2#3:#4\\{\def\epsf@testit{#1#2}
                        \ifx\epsf@testit\epsf@start\else
\typeout{Warning! File does not start with `\epsf@start'.  It may not be a PostScript file.}
                        \fi
                        \@psfiletrue} 
{\catcode`\%=12\global\let\epsf@percent=
\long\def\epsf@aux#1#2:#3\\{\ifx#1\epsf@percent
   \def\epsf@testit{#2}\ifx\epsf@testit\epsf@bblit
        \@atendfalse
        \epsf@atend #3 . \\%
        \if@atend       
           \if@verbose{
                \typeout{psfig: found `(atend)'; continuing search}
           }\fi
        \else
        \epsf@grab #3 . . . \\%
        \not@eoffalse
        \global\no@bbfalse
        \fi
   \fi\fi}%
\def\epsf@grab #1 #2 #3 #4 #5\\{%
   \global\def\epsf@llx{#1}\ifx\epsf@llx\empty
      \epsf@grab #2 #3 #4 #5 .\\\else
   \global\def\epsf@lly{#2}%
   \global\def\epsf@urx{#3}\global\def\epsf@ury{#4}\fi}%
\def\epsf@atendlit{(atend)} 
\def\epsf@atend #1 #2 #3\\{%
   \def\epsf@tmp{#1}\ifx\epsf@tmp\empty
      \epsf@atend #2 #3 .\\\else
   \ifx\epsf@tmp\epsf@atendlit\@atendtrue\fi\fi}

\chardef\letter = 11
\chardef\other = 12

\newif \ifdebug 
\newif\ifc@mpute 
\c@mputetrue 

\let\then = \relax
\def\r@dian{pt }
\let\r@dians = \r@dian
\let\dimensionless@nit = \r@dian
\let\dimensionless@nits = \dimensionless@nit
\def\internal@nit{sp }
\let\internal@nits = \internal@nit
\newif\ifstillc@nverging
\def \Mess@ge #1{\ifdebug \then \message {#1} \fi}

{ 
        \catcode `\@ = \letter
        \gdef \nodimen {\expandafter \n@dimen \the \dimen}
        \gdef \term #1 #2 #3%
               {\edef \t@ {\the #1}
                \edef \t@@ {\expandafter \n@dimen \the #2\r@dian}%
                \t@rm {\t@} {\t@@} {#3}%
               }
        \gdef \t@rm #1 #2 #3%
               {{%
                \count 0 = 0
                \dimen 0 = 1 \dimensionless@nit
                \dimen 2 = #2\relax
                \Mess@ge {Calculating term #1 of \nodimen 2}%
                \loop
                \ifnum  \count 0 < #1
                \then   \advance \count 0 by 1
                        \Mess@ge {Iteration \the \count 0 \space}%
                        \Multiply \dimen 0 by {\dimen 2}%
                        \Mess@ge {After multiplication, term = \nodimen 0}%
                        \Divide \dimen 0 by {\count 0}%
                        \Mess@ge {After division, term = \nodimen 0}%
                \repeat
                \Mess@ge {Final value for term #1 of 
                                \nodimen 2 \space is \nodimen 0}%
                \xdef \Term {#3 = \nodimen 0 \r@dians}%
                \aftergroup \Term
               }}
        \catcode `\p = \other
        \catcode `\t = \other
        \gdef \n@dimen #1pt{#1} 
}

\def \Divide #1by #2{\divide #1 by #2} 

\def \Multiply #1by #2
       {{
        \count 0 = #1\relax
        \count 2 = #2\relax
        \count 4 = 65536
        \Mess@ge {Before scaling, count 0 = \the \count 0 \space and
                        count 2 = \the \count 2}%
        \ifnum  \count 0 > 32767 
        \then   \divide \count 0 by 4
                \divide \count 4 by 4
        \else   \ifnum  \count 0 < -32767
                \then   \divide \count 0 by 4
                        \divide \count 4 by 4
                \else
                \fi
        \fi
        \ifnum  \count 2 > 32767 
        \then   \divide \count 2 by 4
                \divide \count 4 by 4
        \else   \ifnum  \count 2 < -32767
                \then   \divide \count 2 by 4
                        \divide \count 4 by 4
                \else
                \fi
        \fi
        \multiply \count 0 by \count 2
        \divide \count 0 by \count 4
        \xdef \product {#1 = \the \count 0 \internal@nits}%
        \aftergroup \product
       }}

\def\r@duce{\ifdim\dimen0 > 90\r@dian \then   
                \multiply\dimen0 by -1
                \advance\dimen0 by 180\r@dian
                \r@duce
            \else \ifdim\dimen0 < -90\r@dian \then  
                \advance\dimen0 by 360\r@dian
                \r@duce
                \fi
            \fi}

\def\Sine#1%
       {{%
        \dimen 0 = #1 \r@dian
        \r@duce
        \ifdim\dimen0 = -90\r@dian \then
           \dimen4 = -1\r@dian
           \c@mputefalse
        \fi
        \ifdim\dimen0 = 90\r@dian \then
           \dimen4 = 1\r@dian
           \c@mputefalse
        \fi
        \ifdim\dimen0 = 0\r@dian \then
           \dimen4 = 0\r@dian
           \c@mputefalse
        \fi
        \ifc@mpute \then
                \divide\dimen0 by 180
                \dimen0=3.141592654\dimen0
                \dimen 2 = 3.1415926535897963\r@dian 
                \divide\dimen 2 by 2 
                \Mess@ge {Sin: calculating Sin of \nodimen 0}%
                \count 0 = 1 
                \dimen 2 = 1 \r@dian 
                \dimen 4 = 0 \r@dian 
                \loop
                        \ifnum  \dimen 2 = 0 
                        \then   \stillc@nvergingfalse 
                        \else   \stillc@nvergingtrue
                        \fi
                        \ifstillc@nverging 
                        \then   \term {\count 0} {\dimen 0} {\dimen 2}%
                                \advance \count 0 by 2
                                \count 2 = \count 0
                                \divide \count 2 by 2
                                \ifodd  \count 2 
                                \then   \advance \dimen 4 by \dimen 2
                                \else   \advance \dimen 4 by -\dimen 2
                                \fi
                \repeat
        \fi             
                        \xdef \sine {\nodimen 4}%
       }}

\def\Cosine#1{\ifx\sine\UnDefined\edef\Savesine{\relax}\else
                             \edef\Savesine{\sine}\fi
        {\dimen0=#1\r@dian\advance\dimen0 by 90\r@dian
         \Sine{\nodimen 0}
         \xdef\cosine{\sine}
         \xdef\sine{\Savesine}}}              

\def\psdraft{
        \def\@psdraft{0}
}
\def\psfull{
        \def\@psdraft{100}
}

\psfull

\newif\if@draftbox
\def\psnodraftbox{
        \@draftboxfalse
}
\def\psdraftbox{
        \@draftboxtrue
}
\@draftboxtrue

\newif\if@prologfile
\newif\if@postlogfile
\def\pssilent{
        \@noisyfalse
}
\def\psnoisy{
        \@noisytrue
}
\psnoisy
\newif\if@bbllx
\newif\if@bblly
\newif\if@bburx
\newif\if@bbury
\newif\if@height
\newif\if@width
\newif\if@rheight
\newif\if@rwidth
\newif\if@angle
\newif\if@clip
\newif\if@verbose
\def\@p@@sclip#1{\@cliptrue}

\def\@p@@sfile#1{\def\@p@sfile{null}%
                \openin1=#1
                \ifeof1\closein1%
                       \openin1=\figurepath#1
                        \ifeof1\typeout{Error, File #1 not found}
                           \if@bbllx\if@bblly\if@bburx\if@bbury
                              \def\@p@sfile{#1}%
                           \fi\fi\fi\fi
                        \else\closein1
                            \edef\@p@sfile{\figurepath#1}%
                        \fi%
                 \else\closein1%
                       \def\@p@sfile{#1}%
                 \fi}
\def\@p@@sfigure#1{\def\@p@sfile{null}%
                \openin1=#1
                \ifeof1\closein1%
                       \openin1=\figurepath#1
                        \ifeof1\typeout{Error, File #1 not found}
                           \if@bbllx\if@bblly\if@bburx\if@bbury
                              \def\@p@sfile{#1}%
                           \fi\fi\fi\fi
                        \else\closein1
                            \def\@p@sfile{\figurepath#1}%
                        \fi%
                 \else\closein1%
                       \def\@p@sfile{#1}%
                 \fi}

\def\@p@@sbbllx#1{
                \@bbllxtrue
                \dimen100=#1
                \edef\@p@sbbllx{\number\dimen100}
}
\def\@p@@sbblly#1{
                \@bbllytrue
                \dimen100=#1
                \edef\@p@sbblly{\number\dimen100}
}
\def\@p@@sbburx#1{
                \@bburxtrue
                \dimen100=#1
                \edef\@p@sbburx{\number\dimen100}
}
\def\@p@@sbbury#1{
                \@bburytrue
                \dimen100=#1
                \edef\@p@sbbury{\number\dimen100}
}
\def\@p@@sheight#1{
                \@heighttrue
                \dimen100=#1
                \edef\@p@sheight{\number\dimen100}
}
\def\@p@@swidth#1{
                \@widthtrue
                \dimen100=#1
                \edef\@p@swidth{\number\dimen100}
}
\def\@p@@srheight#1{
                \@rheighttrue
                \dimen100=#1
                \edef\@p@srheight{\number\dimen100}
}
\def\@p@@srwidth#1{
                \@rwidthtrue
                \dimen100=#1
                \edef\@p@srwidth{\number\dimen100}
}
\def\@p@@sangle#1{
                \@angletrue
                \edef\@p@sangle{#1} 
}
\def\@p@@ssilent#1{ 
                \@verbosefalse
}
\def\@p@@sprolog#1{\@prologfiletrue\def\@prologfileval{#1}}
\def\@p@@spostlog#1{\@postlogfiletrue\def\@postlogfileval{#1}}
\def\@cs@name#1{\csname #1\endcsname}
\def\@setparms#1=#2,{\@cs@name{@p@@s#1}{#2}}
\def\ps@init@parms{
                \@bbllxfalse \@bbllyfalse
                \@bburxfalse \@bburyfalse
                \@heightfalse \@widthfalse
                \@rheightfalse \@rwidthfalse
                \def\@p@sbbllx{}\def\@p@sbblly{}
                \def\@p@sbburx{}\def\@p@sbbury{}
                \def\@p@sheight{}\def\@p@swidth{}
                \def\@p@srheight{}\def\@p@srwidth{}
                \def\@p@sangle{0}
                \def\@p@sfile{}
                \def\@p@scost{10}
                \def\@sc{}
                \@prologfilefalse
                \@postlogfilefalse
                \@clipfalse
                \if@noisy
                        \@verbosetrue
                \else
                        \@verbosefalse
                \fi
}
\def\parse@ps@parms#1{
                \@psdo\@psfiga:=#1\do
                   {\expandafter\@setparms\@psfiga,}}
\newif\ifno@bb
\def\bb@missing{
        \if@verbose{
                \typeout{psfig: searching \@p@sfile \space  for bounding box}
        }\fi
        \no@bbtrue
        \epsf@getbb{\@p@sfile}
        \ifno@bb \else \bb@cull\epsf@llx\epsf@lly\epsf@urx\epsf@ury\fi
}       
\def\bb@cull#1#2#3#4{
        \dimen100=#1 bp\edef\@p@sbbllx{\number\dimen100}
        \dimen100=#2 bp\edef\@p@sbblly{\number\dimen100}
        \dimen100=#3 bp\edef\@p@sbburx{\number\dimen100}
        \dimen100=#4 bp\edef\@p@sbbury{\number\dimen100}
        \no@bbfalse
}
\newdimen\p@intvaluex
\newdimen\p@intvaluey
\def\rotate@#1#2{{\dimen0=#1 sp\dimen1=#2 sp
                  \global\p@intvaluex=\cosine\dimen0
                  \dimen3=\sine\dimen1
                  \global\advance\p@intvaluex by -\dimen3
                  \global\p@intvaluey=\sine\dimen0
                  \dimen3=\cosine\dimen1
                  \global\advance\p@intvaluey by \dimen3
                  }}
\def\compute@bb{
                \no@bbfalse
                \if@bbllx \else \no@bbtrue \fi
                \if@bblly \else \no@bbtrue \fi
                \if@bburx \else \no@bbtrue \fi
                \if@bbury \else \no@bbtrue \fi
                \ifno@bb \bb@missing \fi
                \ifno@bb \typeout{FATAL ERROR: no bb supplied or found}
                        \no-bb-error
                \fi
                \if@angle 
                        \Sine{\@p@sangle}\Cosine{\@p@sangle}
                        {\dimen100=\maxdimen\xdef\r@p@sbbllx{\number\dimen100}
                                            \xdef\r@p@sbblly{\number\dimen100}
                                            \xdef\r@p@sbburx{-\number\dimen100}
                                            \xdef\r@p@sbbury{-\number\dimen100}}
                        \def\minmaxtest{
                           \ifnum\number\p@intvaluex<\r@p@sbbllx
                              \xdef\r@p@sbbllx{\number\p@intvaluex}\fi
                           \ifnum\number\p@intvaluex>\r@p@sbburx
                              \xdef\r@p@sbburx{\number\p@intvaluex}\fi
                           \ifnum\number\p@intvaluey<\r@p@sbblly
                              \xdef\r@p@sbblly{\number\p@intvaluey}\fi
                           \ifnum\number\p@intvaluey>\r@p@sbbury
                              \xdef\r@p@sbbury{\number\p@intvaluey}\fi
                           }
                        \rotate@{\@p@sbbllx}{\@p@sbblly}
                        \minmaxtest
                        \rotate@{\@p@sbbllx}{\@p@sbbury}
                        \minmaxtest
                        \rotate@{\@p@sbburx}{\@p@sbblly}
                        \minmaxtest
                        \rotate@{\@p@sbburx}{\@p@sbbury}
                        \minmaxtest
                        \edef\@p@sbbllx{\r@p@sbbllx}\edef\@p@sbblly{\r@p@sbblly}
                        \edef\@p@sbburx{\r@p@sbburx}\edef\@p@sbbury{\r@p@sbbury}
                \fi
                \count203=\@p@sbburx
                \count204=\@p@sbbury
                \advance\count203 by -\@p@sbbllx
                \advance\count204 by -\@p@sbblly
                \edef\@bbw{\number\count203}
                \edef\@bbh{\number\count204}
}
\def\in@hundreds#1#2#3{\count240=#2 \count241=#3
                     \count100=\count240        
                     \divide\count100 by \count241
                     \count101=\count100
                     \multiply\count101 by \count241
                     \advance\count240 by -\count101
                     \multiply\count240 by 10
                     \count101=\count240        
                     \divide\count101 by \count241
                     \count102=\count101
                     \multiply\count102 by \count241
                     \advance\count240 by -\count102
                     \multiply\count240 by 10
                     \count102=\count240        
                     \divide\count102 by \count241
                     \count200=#1\count205=0
                     \count201=\count200
                        \multiply\count201 by \count100
                        \advance\count205 by \count201
                     \count201=\count200
                        \divide\count201 by 10
                        \multiply\count201 by \count101
                        \advance\count205 by \count201
                     \count201=\count200
                        \divide\count201 by 100
                        \multiply\count201 by \count102
                        \advance\count205 by \count201
                     \edef\@result{\number\count205}
}
\def\compute@wfromh{
                \in@hundreds{\@p@sheight}{\@bbw}{\@bbh}
                \edef\@p@swidth{\@result}
}
\def\compute@hfromw{
                \in@hundreds{\@p@swidth}{\@bbh}{\@bbw}
                \edef\@p@sheight{\@result}
}
\def\compute@handw{
                \if@height 
                        \if@width
                        \else
                                \compute@wfromh
                        \fi
                \else 
                        \if@width
                                \compute@hfromw
                        \else
                                \edef\@p@sheight{\@bbh}
                                \edef\@p@swidth{\@bbw}
                        \fi
                \fi
}
\def\compute@resv{
                \if@rheight \else \edef\@p@srheight{\@p@sheight} \fi
                \if@rwidth \else \edef\@p@srwidth{\@p@swidth} \fi
}
\def\compute@sizes{
        \compute@bb
        \compute@handw
        \compute@resv
}
\def\psfig#1{\vbox {
        \ps@init@parms
        \parse@ps@parms{#1}
        \compute@sizes
        \ifnum\@p@scost<\@psdraft{
                \if@verbose{
                        \typeout{psfig: including \@p@sfile \space }
                }\fi
                \special{ps::[begin]    \@p@swidth \space \@p@sheight \space
                                \@p@sbbllx \space \@p@sbblly \space
                                \@p@sbburx \space \@p@sbbury \space
                                startTexFig \space }
                \if@angle
                        \special {ps:: \@p@sangle \space rotate \space} 
                \fi
                \if@clip{
                        \if@verbose{
                                \typeout{(clip)}
                        }\fi
                        \special{ps:: doclip \space }
                }\fi
                \if@prologfile
                    \special{ps: plotfile \@prologfileval \space } \fi
                \special{ps: plotfile \@p@sfile \space }
                \if@postlogfile
                    \special{ps: plotfile \@postlogfileval \space } \fi
                \special{ps::[end] endTexFig \space }
                \vbox to \@p@srheight true sp{
                        \hbox to \@p@srwidth true sp{
                                \hss
                        }
                \vss
                }
        }\else{
                \if@draftbox{           
                        \hbox{\fbox{\vbox to \@p@srheight true sp{
                        \vss
                        \hbox to \@p@srwidth true sp{ \hss \@p@sfile \hss }
                        \vss
                        }}}
                }\else{
                        \vbox to \@p@srheight true sp{
                        \vss
                        \hbox to \@p@srwidth true sp{\hss}
                        \vss
                        }
                }\fi

        }\fi
}}
\def\psglobal{\typeout{psfig: PSGLOBAL is OBSOLETE; use psprint -m instead}}
\psfigRestoreAt
\hyphenation{re-weight-ed}
\newcommand{\mlpage}[3]{\begin{minipage} [#1] {#2}#3 \end{minipage}}
\newcommand{\uni}[1]{\, {\rm #1}}
\newcommand{\ome}{$\omega$ }
\newcommand{\pom}{{\rm I\! P}}
\def\3{\ss}
\addtolength{\textwidth}{1.6cm}
\addtolength{\oddsidemargin}{-.8cm}
\addtolength{\evensidemargin}{-1.0cm}
\begin{document}
\pagestyle{empty}
\begin{titlepage}
\title{ 
\vspace{3cm}
{\bf  Measurement of Elastic {\boldmath \ome}Photoproduction at \mbox{HERA} } 
}
\author{\rm \mbox{ZEUS} Collaboration\\}
\date{   }
\maketitle
\vspace{1.0cm}
\begin{abstract}
The reaction $\gamma p \rightarrow \omega p$
$(\omega \rightarrow \pi^+\pi^-\pi^0$ and
$\pi^0\rightarrow\gamma\gamma)$
has been studied in $ep$ interactions
using the \mbox{ZEUS} detector
at photon-proton centre-of-mass energies 
between $70$ and $90\uni{GeV}$ and $|t| < 0.6\uni{GeV}^2$, 
where $t$ is the squared four momentum transferred at the proton
vertex.
The elastic \ome photoproduction cross section has 
been measured to be
$\sigma_{\gamma p\rightarrow \omega p} = 1.21\pm 0.12\pm 0.23 \,\mu\mbox{b}$.
The differential cross section $d\sigma_{\gamma p\rightarrow \omega p} /d|t|$ 
has an exponential shape $\mbox{e}^{-b |t|}$ with a slope
$b = 10.0\pm 1.2\pm 1.3\uni{GeV}^{-2}$. The
angular distributions of the decay pions are consistent
with {\it s}-channel helicity conservation.
When compared to low energy data, the features of $\omega$ photoproduction
as measured at HERA energies are in agreement with those of a 
soft diffractive process. Previous measurements of the
$\rho^0$ and $\phi$ photoproduction cross sections at HERA show a similar 
behaviour.

\end{abstract}
\thispagestyle{empty}
\end{titlepage}
\newpage

\pagestyle{plain}
\pagenumbering{Roman}                                                         
\hsize=1.045\textwidth
{\footnotesize                                                                
                                                                              
\parindent0.cm                                                                
\parskip0.3cm plus0.05cm minus0.05cm                                          
\begin{center}                                                                
{                      \Large  The ZEUS Collaboration              }          
\end{center}                                                                  
  M.~Derrick,                                                                 
  D.~Krakauer,                                                                
  S.~Magill,                                                                  
  D.~Mikunas,                                                                 
  B.~Musgrave,                                                                
  J.R.~Okrasi\'{n}ski,                                                        
  J.~Repond,                                                                  
  R.~Stanek,                                                                  
  R.L.~Talaga,                                                                
  H.~Zhang  \\                                                                
 {\it Argonne National Laboratory, Argonne, IL, USA}~$^{p}$
% ...
\newpage                                                                      
% ...
$^{  19}$ now at ATLAS Collaboration, Univ. of Munich \\                      
% ...
                                                           %
                                                           %
\newpage   
                                                           %
                                                           %
\begin{tabular}[h]{rp{14cm}}                                                  
$^{a}$ &  supported by the Natural Sciences and Engineering Research          
          Council of Canada (NSERC)  \\                                       
$^{b}$ &  supported by the FCAR of Qu\'ebec, Canada  \\                       
$^{c}$ &  supported by the German Federal Ministry for Education and          
          Science, Research and Technology (BMBF), under contract             
          numbers 057BN19P, 057FR19P, 057HH19P, 057HH29P, 057SI75I \\         
$^{d}$ &  supported by the MINERVA Gesellschaft f\"ur Forschung GmbH,         
          the Israel Academy of Science and the U.S.-Israel Binational        
          Science Foundation \\                                               
$^{e}$ &  supported by the German Israeli Foundation, and                     
          by the Israel Academy of Science  \\                                
$^{f}$ &  supported by the Italian National Institute for Nuclear Physics     
          (INFN) \\                                                           
$^{g}$ &  supported by the Japanese Ministry of Education, Science and        
          Culture (the Monbusho) and its grants for Scientific Research \\    
$^{h}$ &  supported by the Korean Ministry of Education and Korea Science     
          and Engineering Foundation  \\                                      
$^{i}$ &  supported by the Netherlands Foundation for Research on             
          Matter (FOM) \\                                                     
$^{j}$ &  supported by the Polish State Committee for Scientific              
          Research, grants No.~115/E-343/SPUB/P03/109/95, 2P03B 244           
          08p02, p03, p04 and p05, and the Foundation for Polish-German       
          Collaboration (proj. No. 506/92) \\                                 
$^{k}$ &  supported by the Polish State Committee for Scientific              
          Research (grant No. 2 P03B 083 08) and Foundation for               
          Polish-German Collaboration  \\                                     
$^{l}$ &  partially supported by the German Federal Ministry for              
          Education and Science, Research and Technology (BMBF)  \\           
$^{m}$ &  supported by the German Federal Ministry for Education and          
          Science, Research and Technology (BMBF), and the Fund of            
          Fundamental Research of Russian Ministry of Science and             
          Education and by INTAS-Grant No. 93-63 \\                           
$^{n}$ &  supported by the Spanish Ministry of Education                      
          and Science through funds provided by CICYT \\                      
$^{o}$ &  supported by the Particle Physics and                               
          Astronomy Research Council \\                                       
$^{p}$ &  supported by the US Department of Energy \\                         
$^{q}$ &  supported by the US National Science Foundation \\                  
\end{tabular}                                                                 
}                                                                             
                                                           %
\newpage                                                           %

\hsize=\textwidth
\vsize=\textheight
\parindent17.6pt
\parskip0.0pt plus 1.0pt
\pagenumbering{arabic}
\setcounter{page}{1}
\section{Introduction}

Elastic photoproduction of vector mesons, $\gamma p\to Vp$,
has been extensively studied in 
fixed target experiments at centre-of-mass energies up to 
$W \approx 20\uni{GeV}$.
The production of $\rho^0$, \ome and $\phi$
is usually described in the framework of the
vector meson dominance model (VDM)~\cite{kn:VDM}
and Regge theory~\cite{kn:regge}. The $W$ dependence
of the cross section can be parametrised in Regge theory
by the sum of two terms, one due to Pomeron
exchange and the other to Reggeon exchange. 
While the latter falls with $W$, the former is
almost flat. Whereas $\rho^0$ and $\phi$ production
is predominantly due to Pomeron exchange at all energies,
the energy behaviour of \ome production investigated
before HERA suggests a non-negligible contribution from Reggeon exchange.
It is therefore of interest to analyse
$\omega$ photoproduction at HERA, where Pomeron 
exchange should dominate.

More specifically, it is important to establish if the 
features typical of elastic $\rho^0$ and $\phi$ 
vector meson production are also
observed in \ome photoproduction at high energy. Among these features are
the weak dependence of the elastic cross section on $W$,
the exponential shape of the differential 
cross section in $t$, where $t$ is the squared four momentum transferred 
at the proton vertex, and the observation that the vector meson
retains the helicity of the photon ({\it s}-channel helicity conservation, 
SCHC). In addition, a comparison of the photoproduction 
cross sections of the light vector mesons
$\rho^0$~\cite{kn:rhopap,kn:H1rho}, 
\ome and $\phi$~\cite{kn:phipap} at HERA energies
allows another check of their diffractive production mechanism.

This paper reports a measurement of 
the photoproduction of \ome mesons using the
reaction $ep\rightarrow e\omega p$ with the ZEUS detector at HERA.
The \ome meson is observed via its decay
into $\pi^+\pi^-\pi^0(\pi^0\rightarrow \gamma\gamma )$
in the kinematic range $70 < W < 90\uni{GeV}$ and 
$p_T^2 < 0.6\uni{GeV}^2$, where $p_T$ is the transverse momentum of
the \ome with respect to the beam axis.
For these events the scattered positron was not observed in the
detector, thereby restricting the photon virtuality
$Q^2$ to values smaller than $4\uni{GeV}^2$,
with a median $Q^2$ of about $10^{-4}\uni{GeV}^2$.

\section{Kinematics}
\label{sec:kinema}

\begin{figure}[p]
\centerline{\psfig{figure=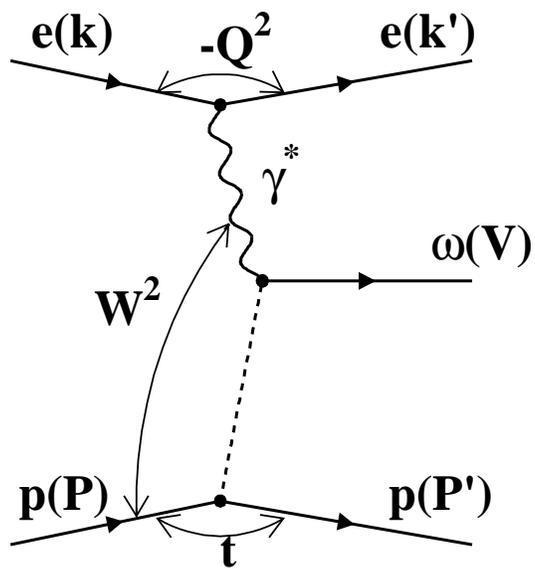,width=\textwidth}}
\caption[Schematic diagram of reaction]{Schematic diagram of elastic \ome 
production in $ep$ interactions. 
} \label{fig:scediag}
\end{figure}

Elastic \ome photoproduction at \mbox{HERA} is measured via the reaction:
$$e(k) + p(P)\rightarrow e(k') + \omega(V) + p(P')\, ,$$
which is shown in Fig.~\ref{fig:scediag}. The symbols in brackets 
denote the four momentum of each particle.

At fixed $ep$ centre-of-mass energy, the inclusive scattering 
of unpolarized positrons and protons can be described by any pair of 
the following variables:
the four momentum squared carried by the photon, $$-Q^2=q^2=(k-k')^2\, ,$$
the centre-of-mass energy squared of the $\gamma^* p$ system,
$$W^2 = (q+P)^2 = -Q^2+2y(k\cdot P)+M_p^2\, ,$$
where $M_p$ is the proton mass, and
the fractional energy transfer of the positron in the proton rest frame, 
   $$y=(q\cdot P)/(k\cdot P)\, .$$

Additional variables are required to describe elastic vector meson 
photoproduction:
the squared four momentum transferred at the proton vertex,
$$t = (P-P')^2 = (q-V)^2 \, ,$$
the angle between the \ome production plane and the positron scattering plane,
and the polar and azimuthal angles of the normal to the \ome decay plane in 
the {\it s}-channel helicity frame.

For the data discussed here, only the decay products of the \ome were
measured. 
The value of $Q^2$ is restricted between the kinematic limit
$Q_{min}^2 = M_e^2\frac{y^2}{1-y}\approx 10^{-9}\uni{GeV}^2$
and $Q^2_{max}\approx 4\uni{GeV}^2$, the latter set by the 
requirement that the scattered positron is not detected in the main detector.
At low values of $Q^2$ the virtual photon is emitted
with negligible transverse momentum
and with longitudinal momentum $p_{Z\gamma}
\approx -E_{\gamma}$, where $E_{\gamma}$ is the photon 
energy\footnote{The ZEUS coordinate system has positive $Z$ in the direction
of flight of the beam protons and the $X$-axis is horizontal, pointing 
towards the centre of \mbox{HERA}. The nominal interaction point is
at $X=Y=Z=0$.}.
Under these assumptions, $t$ and $W$ can be expressed in terms of the
energy $E$ and the longitudinal and transverse momenta
$p_Z$ and $p_T$ of the \ome in the laboratory frame: 
$W^2 \approx 2(E-p_Z)E_p$ and
$t \approx -p_T^2$, where $E_p$ is the proton beam energy.

In order to determine the photoproduction cross section 
$\sigma_{\gamma p\to\omega p}$ from the measured electroproduction cross
section $\sigma_{e p\to e \omega p}$, the following relationship
was assumed~\cite{kn:phipap}, based on VDM~\cite{kn:VDM} and on the one photon 
exchange approximation\footnote{The one photon exchange 
approximation relates the $ep$ cross section to the longitudinal and
the transverse $\gamma^* p$ cross sections for $Q^2\neq 0$, the latter 
being related to $\sigma_{\gamma p\to\omega p}$ by VDM.}:
$$\frac{d^2\sigma_{ep\rightarrow e\omega p}}{dydQ^2} = 
\Phi (y,Q^2)\cdot\sigma_{\gamma p\rightarrow \omega p}(W)
\, ,$$
where
\begin{equation}
\Phi (y,Q^2) = \frac{\alpha}{2\pi Q^2}\left[\frac{1+(1-y)^2}{y}
-\frac{2(1-y)}{y}\left(\frac{Q^2_{min}}{Q^2}-\frac{Q^2}{M_\omega^2}
\right)\right]\left(1+\frac{Q^2}{M_\omega^2}\right)^{-2}  \label{eq:flux}
\end{equation}
is the effective photon flux and $\sigma_{\gamma p\rightarrow \omega p}$
is the photoproduction cross section.

\section{Experimental conditions}
\label{sec:expcon}

\subsection{\mbox{HERA}}
 
During 1994, \mbox{HERA} operated at a proton energy of $820\uni{GeV}$
and a positron energy of $27.5\uni{GeV}$. Typically 153 colliding
positron-proton bunches were stored, along with 17 unpaired
proton and 15 unpaired positron bunches. These additional 
bunches were used to study background from beam-gas interactions.

\subsection{The \mbox{ZEUS} detector}\label{sec:detector}
 
A detailed description of the \mbox{ZEUS} detector can be found 
elsewhere~\cite{kn:status93}.
In addition to
the hadron electron separator, which is described below,
the main components used in this analysis are the same as those
used for the 1994 $\phi$ photoproduction analysis~\cite{kn:phipap}.
Of the latter, only the calorimeter and the tracking chambers are 
mentioned here.

Charged particle momenta are reconstructed from information from
the vertex detector (VXD)~\cite{kn:VXD}, the central 
tracking detector (CTD)\ \cite{kn:CTD} and the rear tracking detector 
(RTD)~\cite{kn:RTD}.
The total angular coverage is $15^\circ < \theta < 170^\circ$, where
$\theta$ is the polar angle in the ZEUS coordinate system.

The high resolution uranium-scintillator calorimeter (CAL)~\cite{kn:CAL} is 
divided into three parts, the forward calorimeter (FCAL), the barrel 
calorimeter (BCAL) and the rear calorimeter (RCAL),
respectively covering the polar angle regions 
$2.6^\circ$ to $36.7^\circ$,
$36.7^\circ$ to $129.1^\circ$, and
$129.1^\circ$ to $176.2^\circ$.
Each part consists of towers which are
longitudinally subdivided into electromagnetic (EMC)
and hadronic (HAC) readout cells. The transverse sizes are
$5\times 20\uni{cm}^2$ for the EMC cells ($10\times 20\uni{cm}^2$ in
the RCAL) and $20\times 20\uni{cm}^2$ for the HAC cells.
From test beam data, energy resolutions with $E$ in GeV of
$\sigma_{E}/E = 0.18/\sqrt{E}$ for electrons and $\sigma_{E}/E = 0.35/\sqrt{E}$
for hadrons have been obtained.

\begin{figure}[p]
  \centerline{
    \psfig{figure=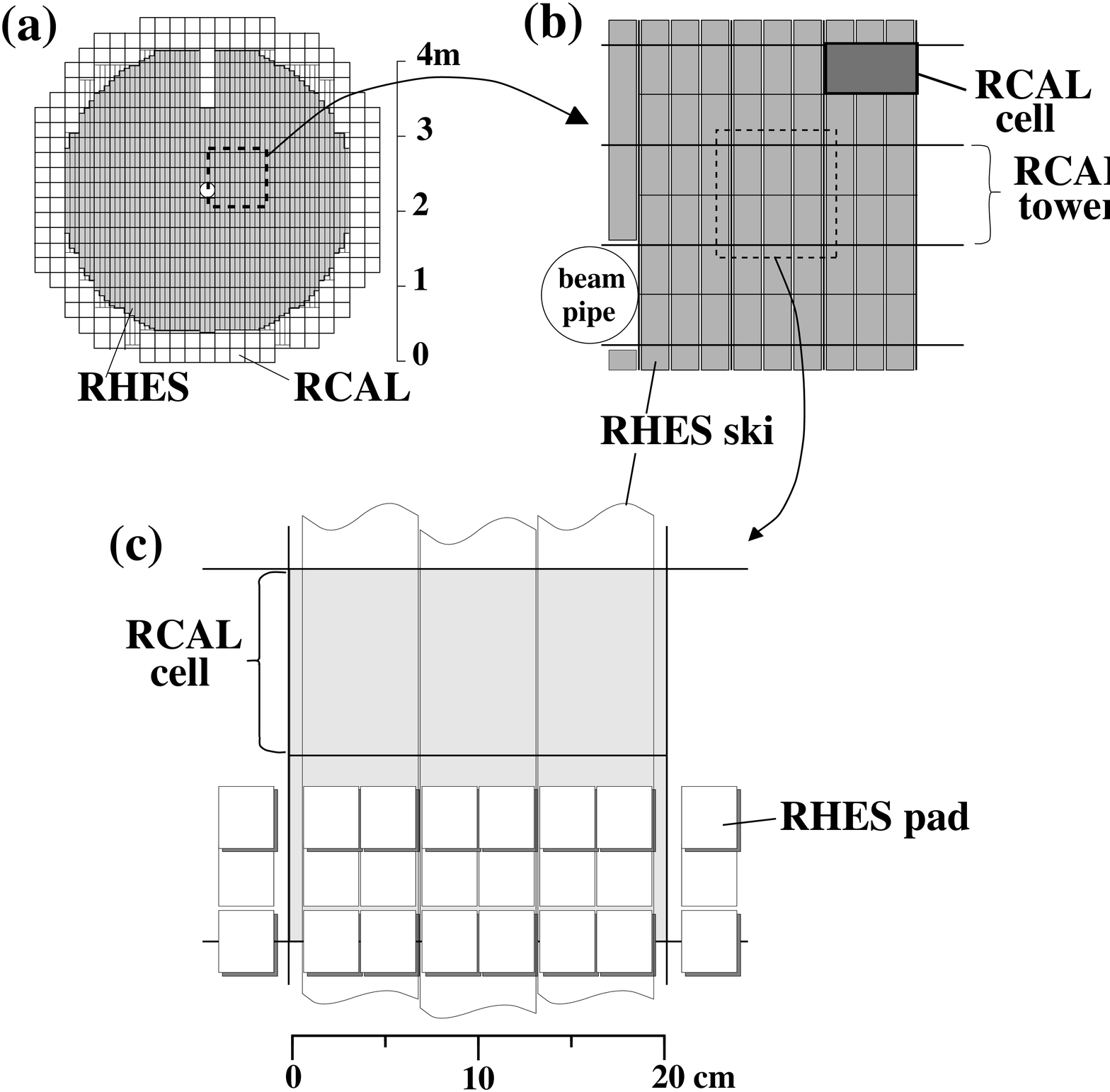,width=.9\textwidth}}
  \caption[Schematic picture of HES]{(a) The area covered by RHES in the
    RCAL as seen from the interaction point. 
    (b) Extended view of (a), indicating the division of the RCAL
    EMC into modules and towers. Two RCAL EMC cells fit into
    one tower. The RHES pads are mounted on support structures, called 
    skis in the figures. (c) Extended view of an RCAL tower. The segmentation
    of RHES into the pads is shown. $6\times 3$ pads fit into
    one RCAL EMC cell.}
  \label{fig:hespic}
\end{figure}

The hadron electron separator (HES)~\cite{kn:HES} 
consists of silicon detectors $400\,\mu{\rm m}$ thick.
In the 1994 running period only the rear part (RHES)
was operational. 
The RHES in the RCAL as seen from the interaction point is shown in
Figure~\ref{fig:hespic}, illustrating the geometrical structure.
The RHES is located in the RCAL at a depth of
$3.3$ radiation lengths, covering an area of about $10\uni{m}^2$.
Each silicon pad has an area of $28.9\times 30.5 \uni{mm}^2$,
providing a spatial resolution of about $9\uni{mm}$ for a single hit pad.
If more than one pad is hit by a shower, a cluster consisting of at most 
$3\times 3$ pads around the most energetic pad is considered. This allows a 
more precise reconstruction of the position, providing a resolution of
about $5\uni{mm}$ for energies greater than $5\uni{GeV}$.
The RHES measures the energy deposited by charged particles near the maximum
of an electromagnetic shower.
This energy is measured in units m.i.p., the energy deposited by a
minimum ionizing particle. 
The mean energy deposit expected for a shower induced by a $1 \uni{GeV}$
photon is 7.6 m.i.p.

\subsection{Trigger}

The conditions of the three level trigger used in this analysis  
are those of
the 1994 $\phi$ photoproduction measurement~\cite{kn:phipap}.

The requirements at the first trigger level consist of a minimum RCAL 
electromagnetic energy deposit of $464\uni{MeV}$ reconstructed by
the calorimeter trigger processor~\cite{kn:CFLT}, a maximum deposit of 
$1250\uni{MeV}$ reconstructed in the FCAL towers surrounding the beam
pipe, and at least one track candidate based on CTD information. 

Upstream proton-gas interactions are rejected at the second and third trigger 
levels using calorimeter time measurements.

The third level trigger also uses the CTD information to reject events
with a reconstructed vertex more than $66\uni{cm}$ away from 
the nominal interaction point along $Z$, with more than four track candidates, 
or with no pair of tracks forming an invariant mass less than 
$1.5\uni{GeV}$ assuming the pion mass for each track.

\section{Event selection and reconstruction}

The data taken during 1994 correspond to a total integrated luminosity
of about $3.2\uni{pb}^{-1}$. After taking the trigger prescaling into account, 
the data presented in this analysis correspond to an effective
integrated luminosity of $894\pm 13\uni{nb}^{-1}$.

\subsection{Selection criteria}\label{sec:evsel}

The final sample of \ome events was selected by imposing the following
offline requirements:
\begin{itemize}
\item Two tracks with opposite charges associated to a common vertex
  and no further  tracks.
\item A well reconstructed $\pi^0$ candidate from  two clusters (as
  defined in section~\ref{sec:pi0rec}) in 
  RCAL and RHES, with at most one additional cluster,
  as described in detail below.
\item No clusters in BCAL or RCAL with energy greater than $200\uni{MeV}$
  and more than $20\uni{cm}$ away from the extrapolated position 
  of either of the two tracks. The cut was not applied to the clusters in the
  RCAL associated to the $\pi^0$ candidate. This cut rejects events
  with additional neutral particles.
\item Transverse momentum of each track greater than $100\uni{MeV}$
  and polar angle $\theta \leq 165^\circ$, to restrict
  the data to a region of good track reconstruction efficiency.
  \item Total energy in FCAL less than $1\uni{GeV}$, in order 
  to limit the contamination by proton dissociative events 
  ($\gamma p\rightarrow \omega N$).
\end{itemize}

\subsection{Reconstruction of the {\boldmath $\pi^0$}}\label{sec:pi0rec}

For the reconstruction of the $\pi^0$ via the decay photon pair, signals in
RCAL and RHES were separately combined into clusters.

RCAL clusters are objects consisting of adjacent calorimeter cells. 
For the present data, clusters were usually formed by one
cell. To reject background from uranium radioactivity,
a minimum cell energy of $100\uni{MeV}$ was required.
This should be compared with the mean
measured photon energy of $500\uni{MeV}$ with a standard deviation of 
$210\uni{MeV}$, which is reproduced by the simulations described below.

RHES clusters consist of at most $3\times 3$ adjacent silicon pads (see 
section~\ref{sec:detector}).
Most (63\%) of the clusters consisted of a single pad.
A cut on the signal from
any RHES pad with less than 1 m.i.p. was applied to reject
noise. The mean RHES signal for this data sample was 4.2 m.i.p. with a 
standard deviation of 2.5 m.i.p.

RHES clusters were assigned to an RCAL 
cluster if they were within a distance of 15 cm (measured in the RHES plane).
RCAL clusters less than 20 cm away from the extrapolated impact point of
a charged track were excluded, thus restricting the sample to clusters 
produced by neutral particles.
Events with exactly two of these neutral RCAL-RHES clusters 
were then selected, allowing at most one additional cluster in RCAL
with no corresponding RHES cluster and an energy of less than $200\uni{MeV}$.
These two RCAL-RHES clusters were required to have an energy deposition in the
electromagnetic part of the calorimeter only.
Fewer than 0.5\% of the RCAL clusters were associated with more than
one RHES cluster.
Even less frequent were events in which the two decay photons were assigned 
to one RHES cluster. 

\begin{figure}[p]
\centerline{\psfig{figure=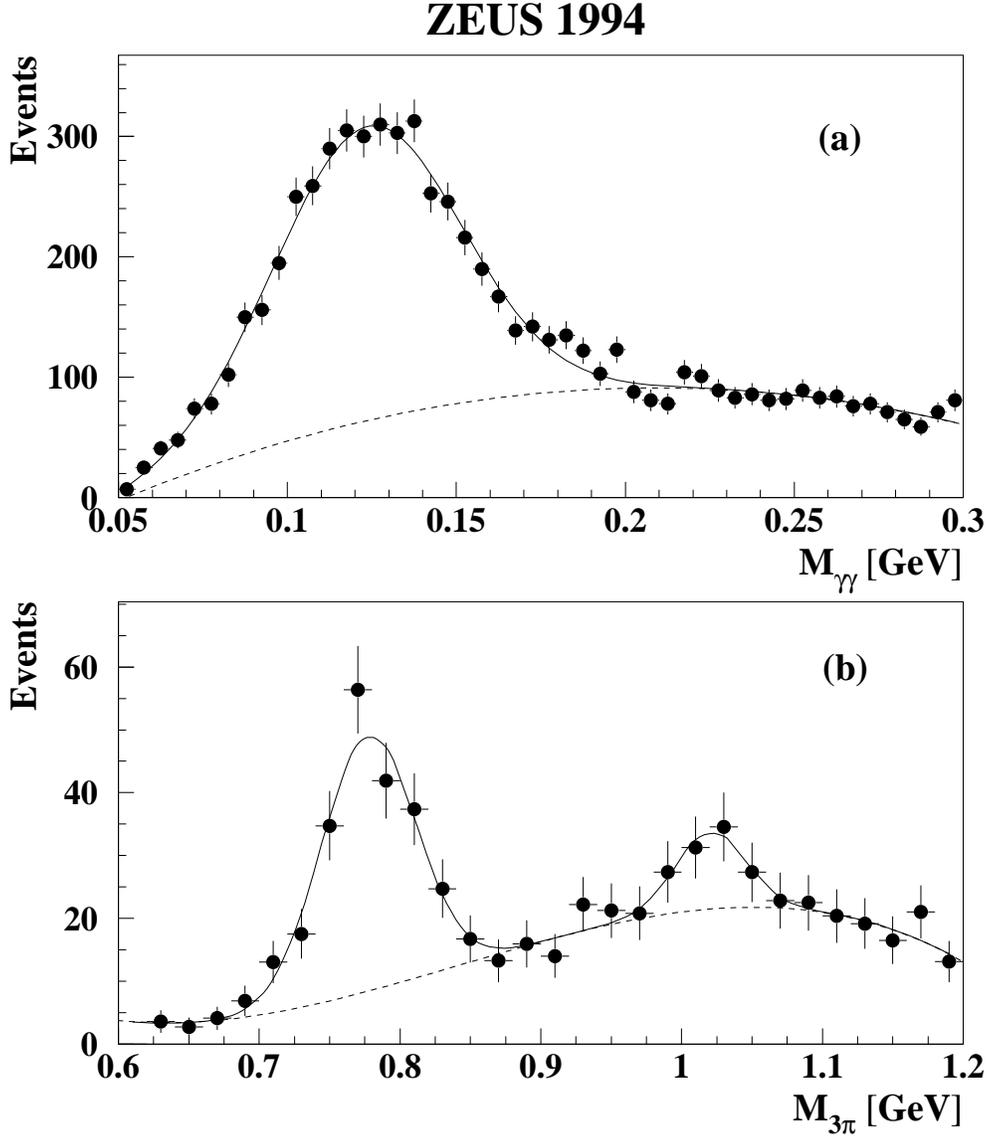,height=.8\textheight}}
\caption[Mass-spectra]{(a) Invariant mass distribution of the two photons.
The full line is the result of the fit explained in the text.
(b) Invariant mass distribution of 
the $\pi^+\pi^-\pi^0$ system after all offline
cuts and the fit based on equation~(\ref{eq:refit}). 
The full line is a fit based on equation~(\ref{eq:mfit}).
In both figures the dashed line indicates the background as determined
by the fits.} \label{fig:mass}
\end{figure}

Using the Monte Carlo simulations described in section~\ref{sec:accs},
the energy of the RCAL clusters was corrected
for losses in the material between the interaction point and the RCAL.
The average correction was approximately 25\%.
The corrected RCAL energies and RHES position information were used to 
calculate the two-cluster invariant mass $M_{\gamma\gamma}$. The
spectrum is plotted in Fig.~\ref{fig:mass}(a). A fit with the sum of a
Gaussian and a second order polynomial yields a mean of the Gaussian of 
$\langle M_{\gamma\gamma}\rangle = 124\pm 1\uni{MeV}$ and a
standard deviation of $28\pm 1\uni{MeV}$.
The difference with respect to the $\pi^0$ mass is a consequence of the 
incomplete description of low energy electromagnetic showers
in the Monte Carlo simulation.
Events with $84 < M_{\gamma\gamma} < 164\uni{MeV}$, i.e. a mass
within $1.5\,\sigma$ of $\langle M_{\gamma\gamma}\rangle$, were selected.

To improve the resolution in the 
four momentum and invariant mass of the $\pi^+\pi^-\pi^0$ system, 
the invariant mass of the two photons was constrained to the $\pi^0$ mass.
Since the $\pi^0$ energies are small, only large opening angles $\alpha$
between the decay photons occur. As the angle $\alpha$ is well determined
due to the precise position measurement of RHES, the resolution in 
$M_{\gamma\gamma}$ is dominated by the energy resolution of RCAL.
Thus only the 
energies were modified in the procedure. The modified values $E_{1_{fit}}$, 
$E_{2_{fit}}$ of the corrected energies $E_1$, $E_2$ of 
the RCAL clusters were determined by minimizing the quantity:
\begin{equation}
\chi^2(E_{1_{fit}},E_{2_{fit}}) = \frac{(E_1 - E_{1_{fit}})^2}{\sigma_{E_1}^2}+
\frac{(E_2 - E_{2_{fit}})^2}{\sigma_{E_2}^2}\, , \label{eq:refit}
\end{equation}
using the constraint:
$$\sqrt{2\cdot E_{1_{fit}}E_{2_{fit}}\cdot (1-\cos \alpha)} = M_{\pi^0}\, ,$$
where $\sigma_{E_i}({\rm GeV})\propto \sqrt{E_i({\rm GeV})}$ 
are the corresponding energy resolutions of the RCAL.

\subsection{Reconstruction of the {\boldmath \ome}}\label{sec:omerec}

To determine the invariant mass of the $\pi^+\pi^-\pi^0$ system and the 
relevant kinematical quantities, the four momentum $p_{3\pi}$ 
of the $\pi^+\pi^-\pi^0$ system was obtained by 
adding up the
momenta of the two tracks, assuming pion masses, and the momentum 
of the $\pi^0$. The latter was determined from the measured RHES positions
and the fitted RCAL energies $E_{1_{fit}}$ and $E_{2_{fit}}$. The quantities 
$W$ and $p_T$ were then derived from $p_{3\pi}$.

The analysis was restricted to the range $70 < W < 90 \uni{GeV}$,
where the acceptance is almost flat. Furthermore the region 
$p_T^2 < 0.6\uni{GeV}^2$ was selected, to limit the background 
contamination due to proton dissociation.

\section{Acceptance calculation and Monte Carlo simulation\label{sec:accs}}

The Monte Carlo generators
PYTHIA~\cite{kn:pythia} and DIPSI~\cite{kn:dipsi} were used
to evaluate the acceptance. 
The former simulates the $\gamma p$ interaction based on VDM and Regge 
theory, while the $Q^2$ spectrum is generated using the ALLM 
parametrisation~\cite{kn:ALLM} of the $ep$ cross section.
The latter uses a model by Ryskin 
\cite{kn:ryskin}, describing vector meson production in terms of 
a fluctuation of the photon into a $q \bar{q}$ pair, 
which interacts with the proton via a Pomeron exchange.
The effective $W$
dependence of the $\gamma p$ cross section for the events generated
was of the type $\sigma \propto W^{0.2}$.
Neither model contains initial or final state radiation.
The effect of radiative corrections on the
cross section has been estimated to be smaller than 4\%~\cite{kn:rhopap}.

The events were generated in the kinematic range
$60\leq W\leq100\uni{GeV}$ and $Q^2_{min}\leq Q^2\leq 4\uni{GeV}^2$.
In order to adjust the Monte Carlo calculation to the data, the 
differential cross section $d\sigma /d|t|$ (see section~\ref{sec:diffcs})
was calculated using the default parameters of PYTHIA
and the Monte Carlo events were then re-weighted with the
measured slope parameter. 
The angular 
distribution of the decay pions was assumed to be that implied by SCHC. 
The reconstructed $W$, $p_T^2$ and decay angular distributions
of the Monte Carlo sample agree well with those of the data,
as do the RCAL and RHES energy distributions of the decay photons.

The RCAL trigger efficiency was 
determined using the data rather than a Monte Carlo simulation. To this 
purpose a sample of charged pions from the decay of $\rho^0$ mesons
produced in elastic photoproduction was used.
Since one of the two pions is sufficient to trigger the event,
the efficiency for RCAL to trigger on a charged pion was evaluated 
as the fraction of events in which
the second pion could have satisfied the trigger and in which it actually did.
The uncertainty in the resulting RCAL trigger efficiency is limited by 
statistics and is 6\%.
The contribution of the photons from the $\pi^0$ decay to the
RCAL trigger decision was determined using the Monte Carlo events
described above.
The total efficiency of the trigger is about 30\%. The largest contribution
to the inefficiency is the RCAL energy threshold (about 50\%).

\begin{figure}[p]
\centerline{\psfig{figure=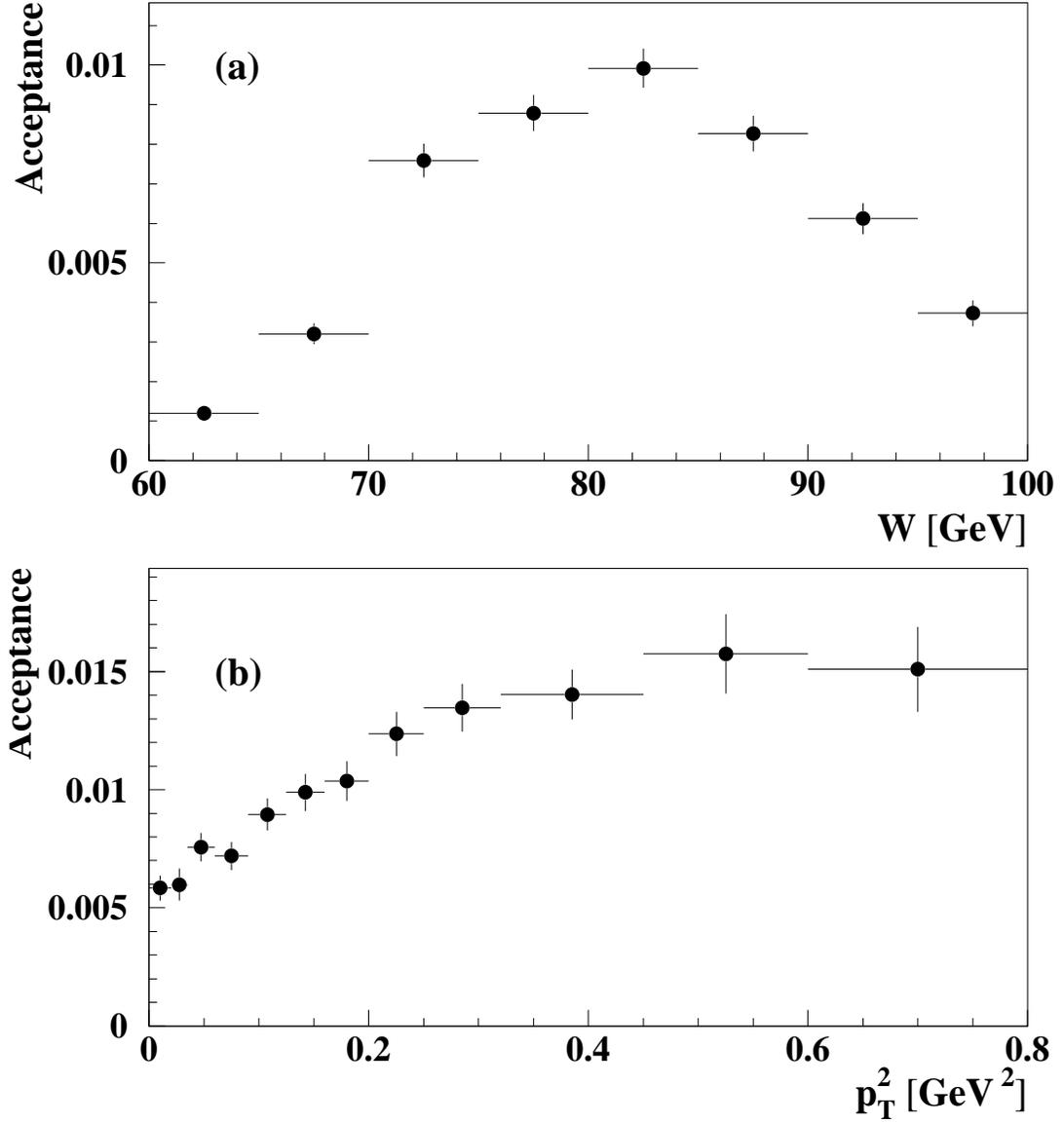,height=.85\textheight}}
\caption[Acceptances]{Acceptance for the process $e p\to e\omega p$.
(a) Acceptance as a function of $W$ in the range $|t|<0.6\uni{GeV}^2$. 
(b) Acceptance as a function of $p_T^2$ in the range $70<W<90\uni{GeV}$.}
\label{fig:accept}
\end{figure}

The acceptance as a function
of $W$ and $p_T^2$ is shown in Figure~\ref{fig:accept}. 
The drop of the acceptance with increasing 
$W$ or with decreasing $p_T^2$ is due to
one or more of the \ome decay products escaping detection in the rear region 
close to the beam pipe. Conversely, with decreasing $W$
the energy of the photons from the $\pi^0$ decay falls below the
value of the cut on the calorimeter energy, thus decreasing the acceptance.

The acceptance evaluated with PYTHIA was used for
the cross section determination.
The average acceptance in the 
region $70 < W < 90\uni{GeV}$ is 0.89\%.
It is dominated by the $\pi^0$ reconstruction, which has itself an efficiency 
of 8\%. The efficiency is further reduced by the 
trigger, the requirement of two tracks (55\%) and the other offline 
requirements (about 70\%).
The model dependence of the acceptance determination was estimated by
changing the distributions of $W$, $t$ and the \ome decay angles in
the Monte Carlo simulation within the 
statistical uncertainties allowed by the data.

The PYTHIA Monte Carlo sample shows agreement between the reconstructed
and generated values of $W$ and $p_T^2$.
The relative resolution in $W$ is 6\% and the resolution in $p_T^2$
is better than $0.04\uni{GeV}^2$.
They are both dominated by the
resolution in the energy measurement of the $\pi^0$ decay photons.

A determination of the elastic cross section 
for $\gamma p\to\phi p$ using the decay $\phi\to\pi^+\pi^-\pi^0$ was
used as a consistency check (see section~\ref{sec:totcs}).
Also in this case the acceptance was determined using
PYTHIA. SCHC was assumed and the events were weighted
according to the measured $t$ distribution for this reaction~\cite{kn:phipap}.
The average acceptance in the region $70 < W < 
90\uni{GeV}$ is 1.5\%.

\section{Background}\label{sec:backgr}

Three sources of background were studied. The first is due to
inelastic reactions with dissociation of the photon into a system of large 
mass. These events produce a smooth mass spectrum, which is parametrised by a
polynomial as discussed in section~\ref{sec:masssp}. A fraction of these
events have an \ome in the final state.
The efficiency for reconstructing only the \ome meson is $10^{-5}$.
Therefore the contribution of this background to the resonance is 
negligible. 

The main source of background is the inelastic
reaction $\gamma p\rightarrow \omega N$, where $N$ is a hadronic system
produced by the dissociation of the proton. 
This background was statistically removed using
the results of~\cite{kn:rhopap}, based on a Monte
Carlo simulation with a cross section of the form 
$d^2\sigma /dt dM_N^2 \propto e^{-b|t|}/M_N^\beta$ ($M_N$ being the mass 
of the state $N$), with $b=4.5$ GeV$^{-2}$ and $\beta=2.5$.
In~\cite{kn:rhopap} the fraction of elastic $\rho^0$ events was
determined as a function of $p_T^2$.
A similar determination using the present data was not
possible due to statistical limitations. 
The same fraction of elastic events was therefore assumed here and a $p_T^2$
dependent weight applied to each event to obtain the elastic \ome cross 
section. The overall
effect is to lower the cross section by $16\pm 9\%$.
The relative uncertainty on the correction was assumed 
to be the same as in~\cite{kn:rhopap}.
The correction was extrapolated to $p_T^2 = 0.6\uni{GeV}^2$
from the range of the measurement presented in~\cite{kn:rhopap}.

Contamination from interactions of the proton or positron beam with the 
residual gas in the beam pipe was estimated from the unpaired bunches 
to be below 2\%.

\section{Results}

\subsection{Analysis of the mass spectrum \label{sec:masssp}}

The spectrum of the invariant mass of the
$\pi^+\pi^-\pi^0$ system $M_{3\pi}$, after all offline cuts and the fit 
constraining the $\gamma\gamma$ mass according to
equation~(\ref{eq:refit}), is shown in Figure~\ref{fig:mass}(b).
In addition to the \ome signal, a second one is visible,
which is due to the elastic photoproduction of the $\phi$ meson,
$\gamma p\rightarrow\phi p$(\mbox{$\phi\rightarrow\pi^+\pi^-\pi^0$}).
The spectrum was fitted with the function:
\begin{equation}
 {\rm f}(M_{3\pi}) = {\rm g}_1(M_{3\pi}) + 
 {\rm g}_2(M_{3\pi}) + \zeta(M_{3\pi})\, , 
\label{eq:mfit}
\end{equation}
where ${\rm g}_1$, ${\rm g}_2$ are convolutions of a 
non-relativistic Breit-Wigner function
with a Gaussian to describe the \ome and $\phi$ resonances.
$\zeta$ is a third order polynomial representing the
background, mainly due to contamination under the $\pi^0$ peak
and to inelastic processes in which a $\pi^0$ is produced.
The fitted values of the \ome and $\phi$ masses are
$778\pm 3\uni{MeV}$ and $1020\pm 9\uni{MeV}$, respectively, compatible
with those of the Particle Data Group (PDG)~\cite{kn:PDG}
and  with Monte Carlo expectations. The fitted values of the Gaussian
standard deviations
are $32\pm 4\uni{MeV}$ and $26\pm 11\uni{MeV}$, respectively, also
in accord with the Monte Carlo;
they are a measure of the resolution of the apparatus.

The number of $\omega$ and $\phi$ candidates observed after background
subtraction was determined by integrating the fitted Breit-Wigner functions
within the kinematic limits. This yields
$N_{\omega} = 172\pm 17$ and $N_\phi = 38\pm 15$.

\subsection{Elastic {\boldmath $\gamma p\rightarrow \omega p$} 
cross section \label{sec:totcs}}

The elastic $\gamma p\rightarrow \omega p$ cross section is given by:
$$\sigma_{\gamma p\rightarrow \omega p} = \frac{N_\omega}{\epsilon\Phi 
{\cal L} B}\, ,$$
where $N_\omega$ is the total number of observed events after the 
statistical subtraction of the inelastic background,
$\epsilon$ is the total acceptance, $\cal L$ is the effective 
integrated luminosity and $\Phi = 0.0203$ is the effective
photon flux as given by equation~(\ref{eq:flux}) after integration over 
the selected $W$ and $Q^2$ ranges.
The branching ratio of the $\omega\rightarrow \pi^+\pi^-\pi^0(\pi^0
\rightarrow\gamma\gamma )$ decay 
is $B=B_{\omega\rightarrow 3\pi}\cdot B_{\pi^0\rightarrow\gamma\gamma} = 
0.877$~\cite{kn:PDG}.
In the region of $|t|<0.6\uni{GeV}^2$, and averaged over the 
range $70 < W <90\uni{GeV}$, the cross section is:
$$\sigma_{\gamma p\rightarrow \omega p} = 1.21\pm 0.12(\rm stat.)
\pm 0.23(\rm syst.)\,\mu{\rm b}\, .$$
The systematic error was obtained by adding in quadrature
the individual contributions listed in Table~\ref{tab:syserr}; the dominant 
ones are from 
the acceptance calculation and the inelastic background subtraction.

\begin{figure}[p]
\centerline{\psfig{figure=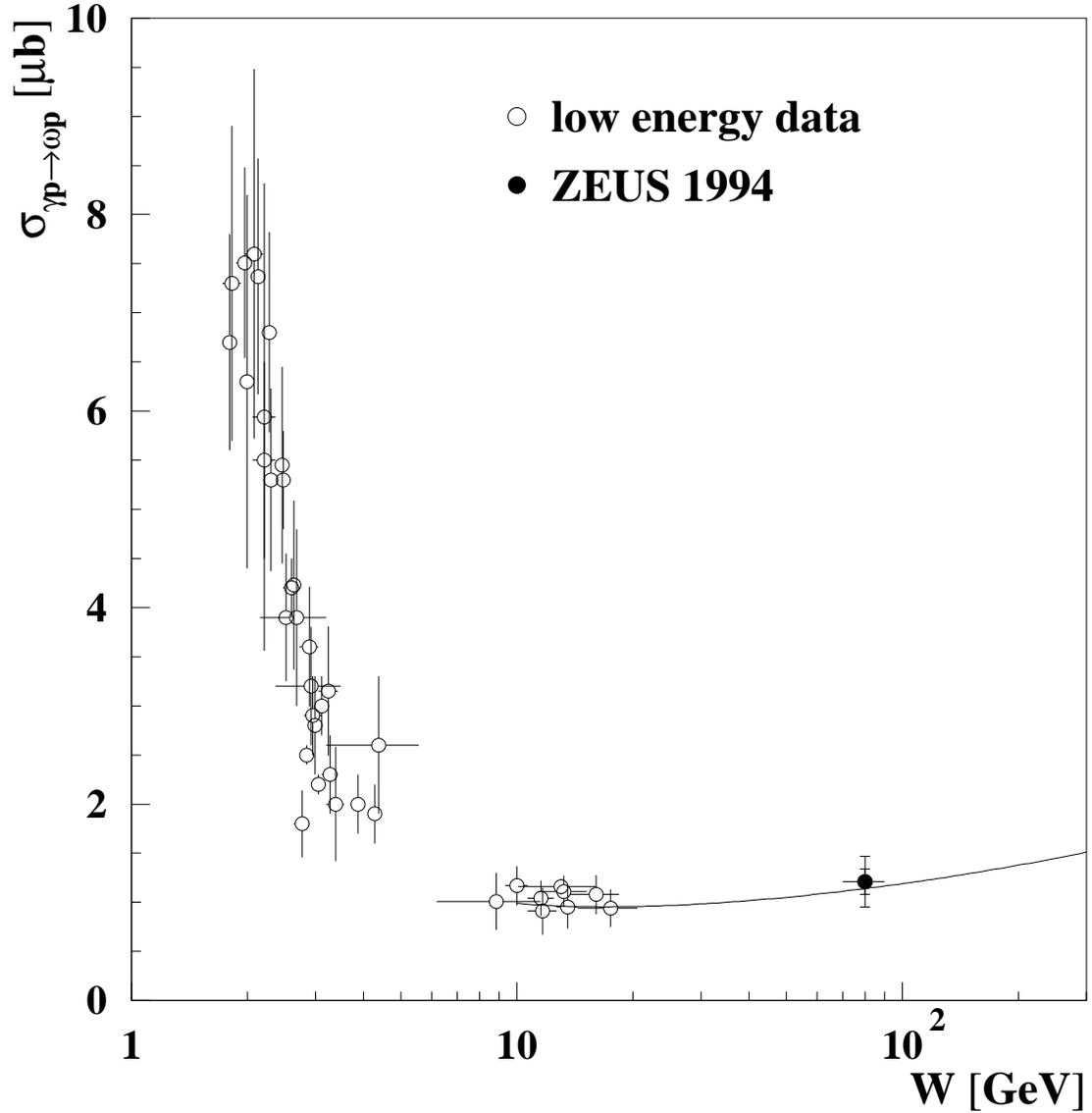,width=.95\textwidth}}
\caption[Elastic cross section]{The elastic $\gamma p \rightarrow \omega p$
cross section measured in this experiment (solid circle) compared with
the results of fixed target experiments~\cite{kn:crouch}-\cite{kn:busen} 
(open circles). For the ZEUS measurement the inner part of the
vertical error bar shows the statistical error, while 
the outer one shows the statistical and systematic errors added in quadrature. 
The horizontal bars indicate the $W$ range covered
by the measurements. The line is a parametrisation 
based on Regge fits to hadronic data~\cite{kn:schuler}.}\label{fig:totcs}
\end{figure}

The resulting elastic $\gamma p\to \omega p$ cross section at an 
average energy of $\langle W\rangle = 80\uni{GeV}$ 
is shown in Figure~\ref{fig:totcs} together with previous results from
fixed target experiments~\cite{kn:crouch}-\cite{kn:busen}.
The $W$ dependence of the cross sections in the range
$10 < W < 100\uni{GeV}$ is found to be weak, as predicted by 
Regge fits to hadronic cross sections~\cite{kn:schuler,kn:donlan}.

As a consistency check, the elastic 
$\gamma p\to \phi p$ cross section 
was determined, using $B=B_{\phi\to 3\pi}\cdot B_{\pi^0\to\gamma\gamma}
=0.154$~\cite{kn:PDG} and $\Phi=0.0207$, resulting in
$\sigma_{\gamma p\to\phi p} = 0.9\pm 0.3(\rm stat.)\,\mu\rm b$.
This value agrees with $\sigma_{\gamma p\to\phi p} 
= 0.96\pm 0.19^{+0.21}_{-0.18}\,\mu\rm b$ determined
using the reaction $\gamma p\to\phi p\;(\phi\to K^+ K^-)$
in the kinematic range $60<W <80 \uni{GeV}$ and 
$|t|<0.5\uni{GeV}^2$~\cite{kn:phipap}.

\subsection{Differential cross section {\boldmath 
$d\sigma_{\gamma p\to\omega p}/d|t|$}}\label{sec:diffcs}

\begin{figure}[p]
\centerline{\psfig{file=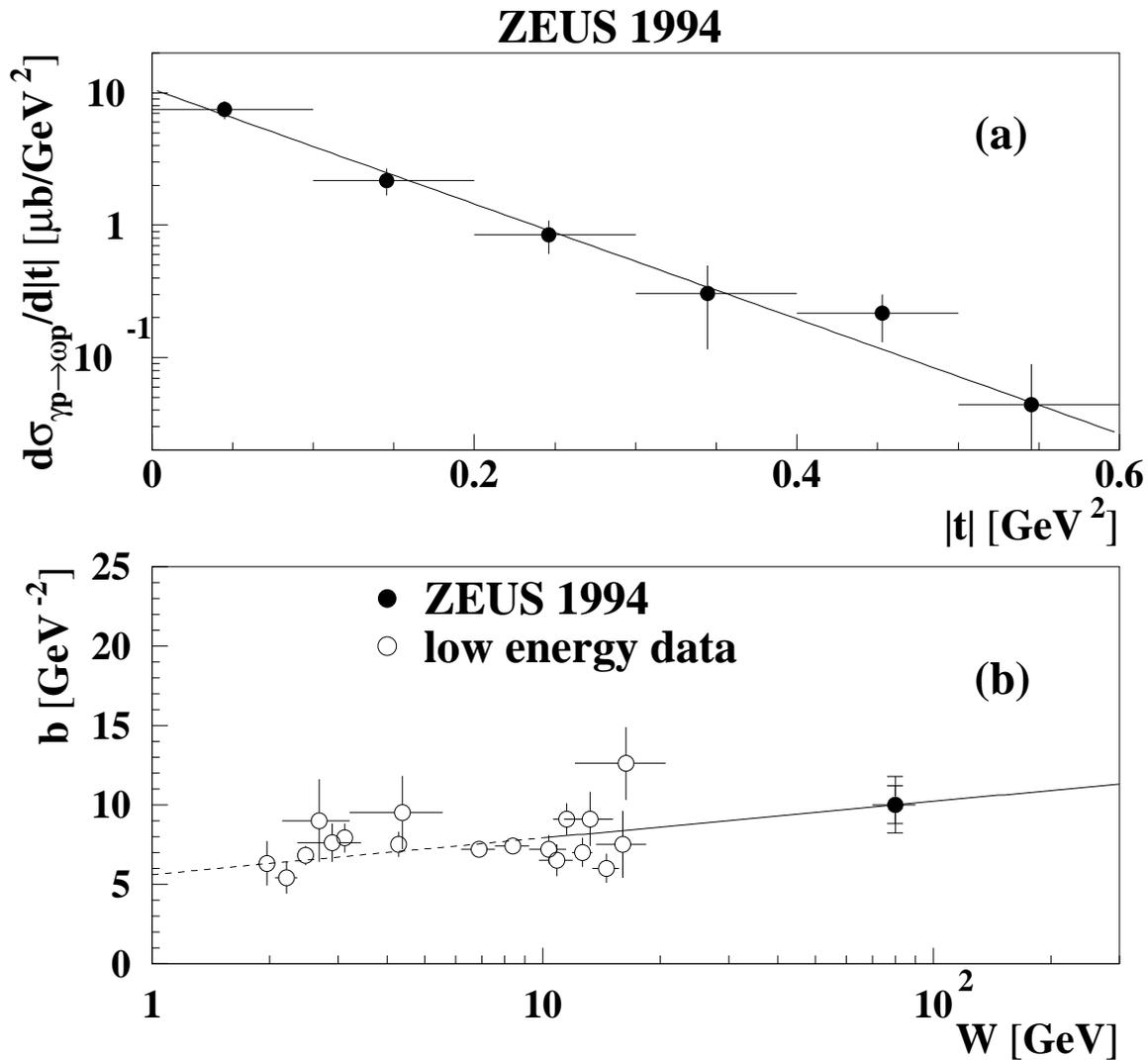,height=.75\textheight}}
\caption[Differential cross section $d\sigma_{\gamma p\to\omega p} /d|t|$]
{(a) The differential cross section $d\sigma_{\gamma p\to\omega p} /d|t|$;
the line shows the result of the fit with functional form~(\ref{eq:difcsfit}).
(b) Exponential slope $b$ of $d\sigma_{\gamma p\to\omega p} /d|t|$
as observed in this experiment (solid circle) compared with low energy 
data~\cite{kn:erbe,kn:davier,kn:ballam,kn:egloff,kn:break,kn:atkinson,kn:busen}
(open circles). 
For the ZEUS measurements the inner part of the vertical error bar shows
the statistical error, while the outer one shows the statistical and systematic
errors added in quadrature.
The horizontal bars indicate the $W$ range covered
by the measurements. The line is given by equation~(\ref{eq:tslopreg}). It was 
constrained to pass through the ZEUS data point.}\label{fig:diftcs}
\end{figure}

In order to derive the differential cross section 
$d\sigma_{\gamma p\to\omega p}/d|t|$,
the fit to the mass spectrum described in section~\ref{sec:masssp}
was repeated in bins of $p_T^2$. 
To reconstruct the $t$ distribution, a bin-by-bin correction~\cite{kn:rhopap},
given by the ratio of the generated $t$ and the reconstructed
$p_T^2$ distribution in the PYTHIA Monte Carlo sample, was used to
compensate for the difference between $t$ and $p_T^2$, which is a consequence
of $Q^2$ not being measured.
The result is shown in Figure~\ref{fig:diftcs}(a).
The data were fitted in the range $0 < |t| < 0.6\uni{GeV}^2$ using the 
functional form:
\begin{equation}
\frac{d\sigma_{\gamma p\to \omega p}}{d|t|} = 
A\cdot{\rm e}^{-b |t|}\, , \label{eq:difcsfit}
\end{equation}
yielding:
\begin{eqnarray}
A & = & 10.7 \pm 2.2(\rm stat.)\pm 2.3(\rm syst.)\,\mu{\rm b}/
\uni{GeV}^2\, , \nonumber\\
b & = & 10.0\pm 1.2(\rm stat.)\pm 1.3(\rm syst.)\uni{GeV}^{-2}\, . 
\nonumber
\end{eqnarray}
The systematic error in $A$ is dominated by the uncertainty on the acceptance
(sensitivity to cuts (16\%) and model dependence (9\%));
the other contributions are the inelastic background subtraction, 
radiative corrections and luminosity, as listed in Table~\ref{tab:syserr}.
The dominant contribution to the
systematic error in $b$ is also the uncertainty on the acceptance
(sensitivity to cuts (12\%) and model dependence (4\%)).

The slope $b$ is compared in Figure~\ref{fig:diftcs}(b)
with the results of previous experiments 
\cite{kn:erbe,kn:davier,kn:ballam,kn:egloff,kn:break,kn:atkinson,kn:busen}. 
From a study of
diffractive hadronic processes, assumed to be mediated by Pomeron exchange,
Regge theory gives the following dependence of the slope $b$ on 
$W$ (with $W$ in GeV)~\cite{kn:regge}:
\begin{equation}
b = b_0 + 2\alpha'_\pom\;\ln W^2\, , \label{eq:tslopreg}
\end{equation}
with $\alpha'_\pom = 0.25 \uni{GeV}^{-2}$. 
The line shown in Figure~\ref{fig:diftcs}(b) represents the dependence
according to equation~(\ref{eq:tslopreg}), with $b_0$ chosen such that
it passes  through the ZEUS point. 
This behaviour is in good agreement with the data points at high
energies ($W>10\uni{GeV}$), where equation~(\ref{eq:tslopreg}) is
expected to hold.

\subsection{Decay angular distributions}\label{sec:decang}

\begin{figure}[p]
\centerline{\psfig{file=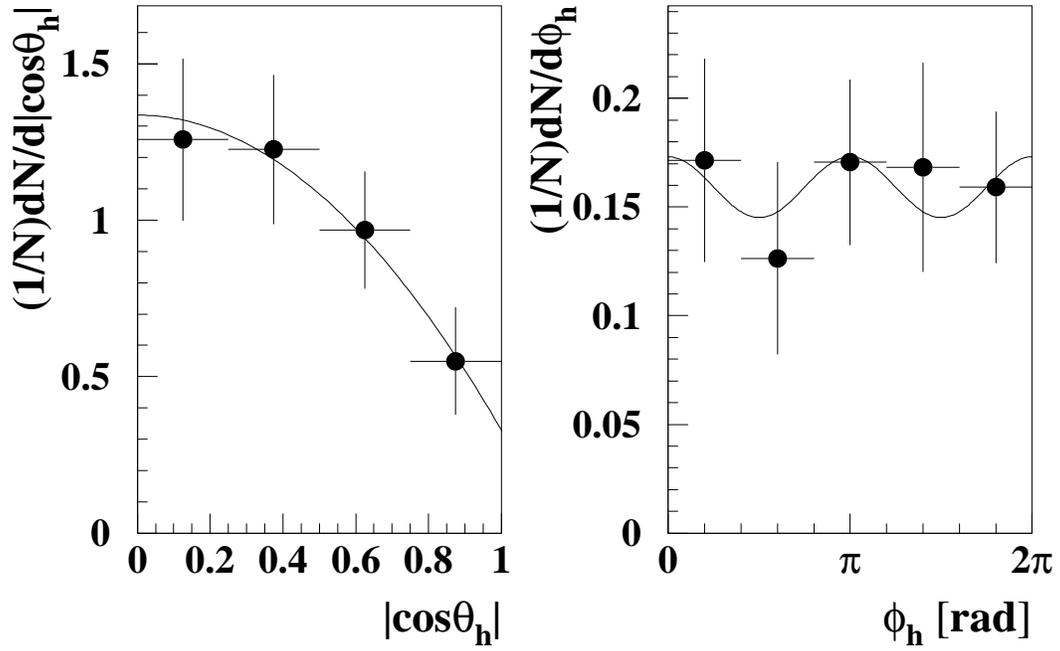,width=.9\textwidth}}
\caption[Corrected $| \cos\theta_h |$ and $\phi$ distributions.]
{Acceptance corrected distribution of $| \cos\theta_h |$ and $\phi_h$, where
$\theta_h$ and $\phi_h$ are the polar and azimuthal angles of the
\ome meson decay plane in the {\it s}-channel helicity frame, respectively. 
Fits according to equations~(\ref{eq:wthe}) and~(\ref{eq:wphi})
are superimposed.}\label{fig:angdist}
\end{figure}

The polar and azimuthal angles $\theta_h$ and $\phi_h$ of the normal
to the \ome meson decay plane in the {\it s}-channel helicity frame
were used to determine elements of
the \ome spin-density matrix~\cite{kn:decang}.
The direction of the normal was defined as that of 
$\vec{\pi}^+\times\vec{\pi}^-$, where $\vec{\pi}^+$ ($\vec{\pi}^-$)
is the three momentum of the positively (negatively) charged pion.
The experimental resolution in $\cos\theta_h$ is about $0.05$
and in $\phi_h$ about $0.8\uni{rad}$.
Upon averaging over $\phi_h$ or $\cos\theta_h$ one finds,
respectively~\cite{kn:decang}:
\begin{equation}
\frac{1}{N}\frac{dN}{d\cos\theta_h} = 
\frac{3}{4} [1-r^{04}_{00}+(3r^{04}_{00}-1)\cos^2\theta_h]\, ,
\label{eq:wthe}
\end{equation}
\begin{equation}
\frac{1}{N}\frac{dN}{d\phi_h} = \frac{1}{2\pi} 
[1-2r^{04}_{1 \mbox{-}\!1}\cos 2\phi_h] \, .
\label{eq:wphi}
\end{equation}
As the distribution of $\cos\theta_h$ was found to be symmetric, 
$|\cos\theta_h|$ was considered instead.
The acceptance corrected distributions of $| \cos\theta_h |$ and $\phi_h$ are
shown in Figure~\ref{fig:angdist}. They were
obtained by repeating the mass fits described in section~\ref{sec:masssp}, 
applied in each bin of $| \cos\theta_h |$ and $\phi_h$.
Fits of the functions~(\ref{eq:wthe}) and~(\ref{eq:wphi}) to the
corrected data yield
$$r^{04}_{00} = 0.11\pm 0.08(\rm stat.)\pm 0.26(\rm syst.)$$
and
$$r^{04}_{1 \mbox{-}\!1} = -0.04\pm 0.08(\rm stat.)\pm 0.12(\rm syst.) \, .$$
Both values are compatible with the prediction 
of {\it s}-channel helicity conservation: $r^{04}_{00} = 0.10$ (assuming a
$Q^2$ dependence of the ratio of the longitudinal to the transverse
$\gamma^* p$ cross section from VDM as also used in 
equation (\ref{eq:flux})) and $r^{04}_{1 \mbox{-}\!1} = 0$.

\section{Comparison of elastic light vector meson photoproduction measurements}

In addition to the measurement presented above,
elastic photoproduction of $\rho^0$ \cite{kn:rhopap,kn:H1rho} and 
$\phi$ \cite{kn:phipap} have also been measured at HERA. 
Table~\ref{tab:sumtcshi} gives a summary of the ZEUS results. 
The elastic $\rho^0$ cross section has also been measured by the H1 
collaboration~\cite{kn:H1rho}, who find a cross section of 
$9.1\pm 0.9\pm 2.5\,\mu{\rm b}$ at $\langle W\rangle = 55\uni{GeV}$
and $13.6\pm 0.8\pm 2.4\,\mu {\rm b}$ at $\langle W\rangle = 187\uni{GeV}$.
We compare in the following our $\rho^0$, \ome and $\phi$ measurements
with each other and with measurements at lower energy,
which are presented in Table~\ref{tab:sumtcsfix}.

From the data of Tables~\ref{tab:sumtcshi} and~\ref{tab:sumtcsfix} 
we calculated the $\rho^0$ to \ome and $\rho^0$ to $\phi$
cross section ratios $\sigma_{\gamma p \to \rho^0 p}/
\sigma_{\gamma p \to \omega p}$ and $\sigma_{\gamma p \to \rho^0 p}/
\sigma_{\gamma p \to \phi p}$ for $\langle W\rangle 
= 12$ or $14\uni{GeV}$ and $70$ or $80\uni{GeV}$.
The results are given in Table~\ref{tab:csrat}.
If there is the same
soft diffractive process mediated by Pomeron exchange for
all three vector mesons, 
no significant energy dependence in these ratios is expected.

Table~\ref{tab:csrate} shows the energy dependence of $\rho^0$, \ome and
$\phi$ elastic photoproduction cross sections, comparing data from this
experiment at $\langle W\rangle =70\uni{GeV}$ or 
$\langle W\rangle=80\uni{GeV}$ with
data from fixed target experiments from Table~\ref{tab:sumtcsfix}.

Using the optical theorem and VDM, the differential cross section
$d\sigma_{\gamma p \to Vp}/d|t|$ for vector meson photoproduction,
with $V=\rho^0$, \ome or $\phi$, can be related to the total vector
meson proton cross section $\sigma_{Vp}$, assuming that the real part
of the amplitude is zero:
\begin{equation}
{\left.\frac{ d\sigma_{\gamma p \to Vp}}{d|t|}~\right|_{t=0}}
             =  \frac{4\pi \alpha}{f_V^2}~%
               \frac{1}{16\pi}\,{\sigma}^2_{Vp}\, ,
\label{eq:opttheo}
\end{equation}
where $f_V^2/4\pi$ is the vector meson photon coupling constant.
Using equation~(\ref{eq:opttheo}) and the approximate relation:
$${\left. \frac{d\sigma_{\gamma p \to Vp}}{d|t|}~\right|_{t=0}}
                                     = b\cdot\sigma_{\gamma p \to Vp}\, ,$$
one finds:
$$ \sigma_{\gamma p \to Vp}
=  \frac{4\pi \alpha}{16\pi f_V^2}~\frac{1}{b}~{\sigma}^2_{Vp}.$$

At these large energies, the total $Vp$ cross sections are expected to
increase with energy~\cite{kn:chwawu}
proportionally to $ W^{2\alpha_\pom (0)-2}$, where
$\alpha_\pom (0)$ is the intercept  of the Pomeron trajectory.
With $ \alpha_\pom (0)$ = 1.08~\cite{kn:donlan}, and assuming the slope 
$b$ to increase with $W$ according to 
equation~(\ref{eq:tslopreg}),
one finds an increase in the elastic vector meson photoproduction
cross sections with energy as shown in Table~\ref{tab:csrate}.
The predictions are consistent with the data.
The curve in Figure~\ref{fig:totcs} includes the contribution from Reggeon 
exchange leading to a smaller predicted increase.

Equation~(\ref{eq:opttheo}) can be used to compute the total \ome proton 
cross section. Using $\frac{d\sigma
}{d|t|}{|}_{t=0}=10.7\,\mu{\rm b}/\uni{GeV}^2$ (see Table~\ref{tab:sumtcshi}) 
and $f_\omega^2 /4\pi = 23.6$~\cite{kn:bauer}, one gets
$$\sigma_{\omega p} = 26.0\pm 2.5{\rm (stat.)}^{+3.0}_{-4.2}{\rm (syst.)
\uni{mb}}\, ,$$
where the systematic error includes the error of 
$\frac{d\sigma
}{d|t|}{|}_{t=0}$ and the estimated 
uncertainty of $f_\omega^2 /4\pi$. This value agrees with the
expected value of $28\uni{mb}$ at $W=80\uni{GeV}$, obtained from a
parametrisation of $\sigma_{\omega p} = 
\frac{1}{2} (\sigma_{\pi^+ p} + \sigma_{\pi^- p})$~\cite{kn:schuler}.

\section{Conclusions}

Elastic \ome photoproduction at $\langle W \rangle = 80\uni{GeV}$
and $|t| < 0.6\uni{GeV}^2$ has been measured using the ZEUS detector. 
The elastic cross section is $\sigma_{\gamma p\rightarrow\omega p} = 
1.21\pm 0.12\pm 0.23 \,\mu{\rm b}$. The exponential slope of the
differential cross section $d\sigma_{\gamma p\rightarrow\omega p}/d|t|$
in this $t$ range 
has been determined to be $10.0\pm 1.2\pm 1.3 \uni{GeV}^{-2}$. 
The angular distributions of the decay pions
are consistent with {\it s}-channel helicity conservation.

When compared to low energy data, the features of $\omega$ photoproduction
as measured at HERA energies are in agreement with those of a 
soft diffractive process. Previous measurements of the
$\rho^0$ and $\phi$ photoproduction cross sections at HERA show a similar 
behaviour.

\section{Acknowledgement}
We thank the DESY Directorate for their strong support and
encouragement. The remarkable achievements of the HERA machine group
were essential for the successful completion of this work and are
gratefully appreciated. 

For the design and construction of the RHES special thanks go to
D.~Cassel, S.~Kasai, A.~Montag, K.U.~P\"osnecker, V.~Radeka and R.~Rau,
who contributed in the early stage of the project. 
The engineering and technical support of the following institutes is 
gratefully acknowledged:
Faculty of Physics and Nuclear Techniques, Cracow;
DESY;
Univ. of Hamburg;
Univ. Aut\'{o}noma Madrid; 
Moscow State Univ.;
Univ. of Tokyo and
Weizmann Institute.



\newpage

\begin{table}[t]
  \centerline{
    \begin{tabular}{|ll|c|l|} \hline
      \multicolumn{2}{|l|}{Source} & Error & Refer to \\ 
      \multicolumn{2}{|l|}{ } & [\%] &  \\ \hline
      \multicolumn{2}{|l|}{Luminosity} & 1.5 &  \\
      Acceptance: & $\bullet$\  
      sensitivity to track selection cuts (modify to & & \\
      & $\:\;$\  $p_T>150\uni{MeV}$ or $\theta<163^\circ$ or 
      $\theta<167^\circ$); & 4 & Section~\ref{sec:evsel}\\
      & $\bullet$\  sensitivity to RCAL and RHES energy cuts & & \\
      & $\:\;$\ (change by 50\% or assume a miscalibration & & \\
      & $\:\;$\ of RHES by 20\%); & 8 & 
      Section~\ref{sec:pi0rec}\\
      & $\bullet$\ 
      sensitivity to $M_{\gamma\gamma}$ range (change cut on  & & \\
      & $\:\;$\ $|M_{\gamma\gamma}-\langle M_{\gamma\gamma}\rangle |$
      by $\pm 10\uni{MeV}$); & 9 & Section~\ref{sec:pi0rec}\\
      & $\bullet$\  sensitivity to $W$ range (change cut values by & & \\
      & $\:\;$\ 6\%); & 5 & Section~\ref{sec:omerec}\\
      & $\bullet$\  trigger efficiency; & 6 & Section~\ref{sec:accs}\\
      & $\bullet$\  model dependence. & 3 & Section~\ref{sec:accs}\\
      \multicolumn{2}{|l|}{Radiative corrections} & 4 & 
      Section~\ref{sec:accs} \\
      \multicolumn{2}{|l|}{Inelastic background subtraction} & 11 & 
      Section~\ref{sec:backgr}\\ \hline
      \multicolumn{2}{|l|}{Total (added in quadrature)} & 19 & \\ \hline
    \end{tabular}}
  \caption[Systematic errors]{Contributions to the
    systematic error on $\sigma_{\gamma p\to\omega p}$.}\label{tab:syserr}
\end{table} 

\begin{table}[htp]
$$
\begin{tabular}{|c||c|c|c|}
\hline
 & $ \gamma p \to \rho^0 p$ & $ \gamma p \to \omega p$ & $ \gamma p \to \phi p$
         \\ \hline
$\langle W\rangle$~[GeV] & 70 & 80 & 70    \\
$\sigma$~[$\mu$b] & 14.7 $\pm$ 0.4 $\pm$ 2.4 
        & 1.21 $\pm$ 0.12 $\pm$ 0.23 & 0.96 $\pm$ 0.19$^{+0.21}_{-0.18}$ \\
$|t|$-range~[GeV$^2$] & $<0.5$ & $<0.6$ & $<0.5$ \\
\omit\vrule\dotfill\vrule&\omit\vrule\dotfill\vrule&\omit\vrule\dotfill
\vrule&\omit\vrule\dotfill\vrule \\
$\frac{d\sigma}{d|t|}{|}_{t=0}$~[$\mu$b/GeV$^2$]  & 139 $\pm$ 6 $\pm$ 26
            & $10.7\pm$ 2.2 $\pm$ 2.3 & 7.2 $\pm$ 2.1 $\pm$ 1.5 \\
 $b$~[GeV$^{-2}$] & 10.4 $\pm$ 0.6 $\pm$ 1.1 & 10.0 $\pm$ 1.2 $\pm$ 1.3
                         & 7.3 $\pm$ 1.0 $\pm$ 0.8  \\
$|t|$-range~[GeV$^2$] & $<0.15$ & $<0.6$ & $0.1<|t|<0.5$ \\ \hline
 ref. & \cite{kn:rhopap}   & this paper & \cite{kn:phipap}   \\
\hline
\end{tabular}
$$
\caption[Summary of HERA]
{Summary of results on elastic light vector meson photoproduction
at HERA as measured by ZEUS in the given $t$~range. The values of 
$\frac{d\sigma}{d|t|}{|}_{t=0}$ and of $b$ have been taken from a fit of
the differential cross section of the form $ \frac{d\sigma}{d|t|} = 
\frac{d\sigma}{d|t|}{|}_{t=0}\cdot~{\rm e}^{-b|t|}$ in the given
$t$~range.}
\label{tab:sumtcshi}
\end{table}

\begin{table}[htp]
\begin{center}
\begin{tabular}{|c||c|c|c|}
\hline
 & $ \gamma p \to \rho^0 p$
            & $ \gamma p \to \omega p$
                         & $ \gamma p \to \phi p$  \\ \hline
 $\ \ \ \ \ \langle W\rangle$~[GeV]\ \ \ \ \  
 & 12         & 14         & 14  \\
 $\sigma$~[$\mu$b] &\ \ \ \ 9.25 $\pm$ 0.44 \ \ \  &\ \ \ \ \ 
 1.07 $\pm$ 0.12\ \ \ \ \ 
     & \ \ \ 0.66 $\pm$ 0.08 \ \ \  \\
 ref. & \cite{kn:aston,kn:egloff2}
     &\cite{kn:egloff,kn:break,kn:atkinson,kn:busen}& \cite{kn:egloff2}   \\
\hline
\end{tabular}
\caption[Summary of fixed target]
{Elastic $ \gamma p \to Vp$ cross sections
from fixed target experiments for $W\geq 9\uni{GeV}$. The values listed
were obtained as weighted means of the cited measurements.}
\label{tab:sumtcsfix}
\end{center}
\end{table}

\begin{table}[htp]
\begin{center}
\begin{tabular}{|c||c|c|}
\hline
$\langle W\rangle$ & 
  $\sigma_{\gamma p \to \rho^0 p}/\sigma_{\gamma p \to \omega p}$ &
  $\sigma_{\gamma p \to \rho^0 p}/\sigma_{\gamma p \to \phi p}$ \\ \hline
\ \ \ 12 or 14~GeV\ \ \  & 8.6 $\pm$ 1.0 & 14.0 $\pm$ 1.8 \\
\ \ \ 70 or 80~GeV\ \ \  &  12.2 $\pm$ 3.0 &  15.3 $\pm$ 4.9 \\ \hline
\end{tabular}
\caption[Ratios 1]{Ratios of vector meson photoproduction cross sections.}
\label{tab:csrat}
\end{center}
\end{table}

\begin{table}[htb]
\begin{center}
\begin{tabular}{|l||c|c|c|}
\hline
                   &               &                &                 \\
& $\frac{\textstyle \sigma^{\langle W\rangle =70}_{\gamma p \to \rho^0 p}}
     {\textstyle \sigma^{\langle W\rangle =12}_{\gamma p \to \rho^0 p}}$
  & $\frac{\textstyle \sigma^{\langle W\rangle =80}_{\gamma p \to \omega p}}
       {\textstyle \sigma^{\langle W\rangle =14}_{\gamma p \to \omega p}}$
     & $\frac{\textstyle \sigma^{\langle W\rangle =70}_{\gamma p \to \phi p}}
          {\textstyle \sigma^{\langle W\rangle =14}_{\gamma p \to \phi p}}$  \\
                 &               &                &                 \\ \hline
 Experiment        & 1.59 $\pm$ 0.28 & 1.12 $\pm$ 0.22 & 1.45 $\pm$ 0.46 \\
 Pomeron exchange  & 1.46            & 1.45            & 1.28            \\
\hline
 $b_0$ (from eq.~(\ref{eq:tslopreg})) & 6.15 & 5.75 & 3.05 \\ \hline
\end{tabular}
\caption[Ratios 2]{Ratio of cross sections at high and low $W$.}
\label{tab:csrate}
\end{center}
\end{table}


\newpage

\newpage

\newpage

\newpage

\newpage

\newpage

\end{document}